\documentclass[a4paper,10pt,twoside]{cpc-hepnp}

\usepackage{multicol}
\usepackage{graphicx}
\usepackage{booktabs}
\usepackage{amssymb,bm,mathrsfs,bbm,amscd}
\usepackage[tbtags]{amsmath}
\usepackage{lastpage}
\usepackage{CJK}

\begin{document}



\title{Charged current quasi elastic scattering of $\bar{\nu}_{\mu}$ off $^{12}$C\thanks{Supported by Department of Science and Technology, New Delhi, India }}

\author{%
      Deepika Grover$^{1}$
\quad Kapil Saraswat$^{1}$
\quad Prashant Shukla$^{2,3}$
\quad Venktesh Singh$^{1;1)}$\email{venkaz@yahoo.com}%
}
\maketitle

\address{%
$^1$ Department of Physics, Institute of Science, Banaras Hindu University, Varanasi 221005, India\\
$^2$ Nuclear Physics Division, Bhabha Atomic Research Centre, Mumbai 400085, India\\
$^3$ Homi Bhabha National Institute, Anushakti Nagar, Mumbai 400094, India\\
}

\begin{abstract}
In this work, we study charged current quasi elastic scattering of $\bar{\nu}_{\mu}$ off nucleon and nucleus using a formalism based on Llewellyn Smith (LS) model. Parameterizations by Galster et al. are used for electric and magnetic Sach's form factors of nucleons. We use Fermi gas model along with Pauli suppression condition to take into account the nuclear effects in anti-neutrino - nucleus QES. We calculate $\bar{\nu}_{\mu} - p$ and $\bar{\nu}_{\mu} - ^{12}$C charged current quasi elastic scattering differential and total cross sections for different values of axial mass $M_{A}$ and compare the results with data from GGM, SKAT, BNL, NOMAD, MINER$\nu$A and MiniBooNE experiments. The present theoretical approach gives an excellent description of differential cross section data. The calculations with axial mass $M_{A} = 0.979$ and $1.05$ GeV are compatible with data from most of the experiments.
\end{abstract}

\begin{keyword}
Cross section, Quasi elastic scattering, Axial mass, Form factors
\end{keyword}

\begin{pacs}
13.15.+g, 25.70.Bc, 96.40.Tv
\end{pacs}

\footnotetext[0]{\hspace*{-3mm}\raisebox{0.3ex}{$\scriptstyle\copyright$}2018
Chinese Physical Society and the Institute of High Energy Physics
of the Chinese Academy of Sciences and the Institute
of Modern Physics of the Chinese Academy of Sciences and IOP Publishing Ltd}%

\begin{multicols}{2}

\section{Introduction}

From their first postulation by Wolfgang Pauli in 1930, to explain the continuous energy spectra in beta decay process, the neutrinos have been a major field of research. Neutrinos exist in three flavors (electron, muon and tau neutrinos) along with their anti-particles called anti-neutrinos. Search for more neutrino flavors called sterile neutrinos is still underway. The standard model of particle physics assumes (anti)neutrinos to be massless, however, several (anti)neutrino oscillation experiments have confirmed small but non zero (anti)neutrino masses~\cite{Ahn:2002up,Aliu:2004sq,Ahn:2006zza,Ashie:2005ik,Takeuchi:2011aa,Adamson:2016tbq,Adamson:2016xxw,Adamson:2017qqn,Adamson:2017gxd,Kumar:2017sdq}. Being neutral particles, (anti)neutrinos undergo only weak interaction, i.e. charged current: via exchange of $W^{+}/W^{-}$ boson and neutral current: via exchange of $Z$ boson, with matter through scattering processes such as quasi elastic scattering (QES), resonance pion production (RES) and deep inelastic scattering (DIS), for a review see Ref.~\cite{Formaggio:2013kya}. In charged current (CC) quasi elastic scattering, an (anti)neutrino interacts with a (proton)neutron producing a corresponding lepton and the (proton)neutron changes to (neutron)proton.
\begin{eqnarray}
\nu_{l} + n \rightarrow l^{-} + p .       \\
\bar{\nu}_{l} + p \rightarrow l^{+} + n .
\end{eqnarray}

Precise knowledge of (anti)neutrino CCQES is crucial to high energy physics experiments studying neutrino oscillations and hence extracting neutrino mass hierarchy, mixing angles etc.~\cite{Ahn:2002up,Aliu:2004sq,Ahn:2006zza,Ashie:2005ik,Takeuchi:2011aa,Adamson:2016tbq,Adamson:2016xxw,Adamson:2017qqn,Adamson:2017gxd,Kumar:2017sdq}. Several experimental efforts such as studies at Gargamelle (GGM)~\cite{Bonetti:1977cs,Armenise:1979zg}, SKAT~\cite{Brunner:1989kw}, Brookhaven National Laboratory (BNL)~\cite{Fanourakis:1980si}, Neutrino Oscillation MAgnetic Detector (NOMAD)~\cite{Lyubushkin:2008pe}, Main INjector ExpeRiment for $\nu$ - A (MINER$\nu$A)~\cite{Fields:2013zhk} and Mini Booster Neutrino Experiment (MiniBooNE)~\cite{Aguilar-Arevalo:2013dva} etc. have been performed to describe the quasi elastic scattering of neutrinos and anti-neutrinos off various nuclear targets. GGM studied quasi elastic reactions of neutrinos and anti-neutrinos on propane along with freon target, SKAT bombarded a wide energy band neutrino/anti-neutrino beam onto heavy freon (CF$_{3}$Br) target, BNL used hydrogen (H$_{2}$) as target, NOMAD executed the studies on carbon, MINER$\nu$A projected an anti-neutrino beam with average energy of 3.5 GeV onto a hydrocarbon target and MiniBooNE recorded the data on mineral oil target. A global analysis of neutrino and anti-neutrino QES differential and total cross sections along with the extraction of axial mass $M_{A}$ is presented in Ref.~\cite{Kuzmin:2007kr}.

In this work, we study charged current anti-neutrino - nucleon and anti-neutrino - nucleus ($^{12}$C) QES. To describe CCQES, we use the Llewellyn Smith (LS) model~\cite{LlewellynSmith:1971uhs} and parameterizations by Galster et al.~\cite{Galster:1971kv} for electric and magnetic Sach's form factors of nucleons. For incorporating the nuclear effects, in case of $\bar{\nu}_{\mu}$ scattering off $^{12}$C, we use the Fermi gas model along with Pauli suppression condition~\cite{LlewellynSmith:1971uhs,Smith:1972xh,Kuzmin:2007kr}. We calculate $\bar{\nu}_{\mu} - p$ and $\bar{\nu}_{\mu} - ^{12}$C CCQES differential and total cross sections for different values of axial mass $M_{A}$ and compare the results with experimental data with the goal of finding the most appropriate $M_{A}$ value. This work does not include contribution from 2N2h (two nucleons two holes) effect in QES.

\section{Formalism for quasi elastic $\bar{\nu} - N$ and $\bar{\nu} - A$ scattering}

The anti-neutrino - nucleon charged current quasi elastic differential cross section for a free nucleon at rest is given by~\cite{LlewellynSmith:1971uhs}:
\begin{eqnarray}
\frac{d\sigma^{free}}{dQ^{2}}&=&\frac{M^{2}_{N} ~G^{2}_{F} ~cos^{2}\theta_{c}}{8\pi E^{2}_{\bar{\nu}}}\Bigg[A(Q^{2}) \nonumber \\
&&+ \frac{B(Q^{2}) ~(s-u)}{M^{2}_{N}} + \frac{C(Q^{2}) ~(s-u)^{2}}{M^{4}_{N}}\Bigg] ,  \label{eq:dsigma_free_nucleon}
\end{eqnarray}
where $M_{N}$ is the nucleon mass, $G_{F} ~(= 1.16 \times 10^{-5}$ GeV$^{-2})$ is the Fermi coupling constant, $cos\theta_{c} ~(= 0.97425)$ is the Cabibbo angle and $E_{\bar{\nu}}$ is the anti-neutrino energy. In terms of the mandelstam variables $s$ and $u$, the relation $s-u = 4M_{N}E_{\bar{\nu}} - Q^{2} - m^{2}_{l}$, where $Q^{2}$ is the square of momentum transfer from anti-neutrino to the outgoing lepton and $m_{l}$ is the mass of the outgoing lepton.

The functions $A(Q^{2}), B(Q^{2})$ and $C(Q^{2})$ are defined as~\cite{LlewellynSmith:1971uhs}:
\begin{eqnarray}
A(Q^{2})&=&\frac{(m^{2}_{l} + Q^{2})}{M^{2}_{N}} \Bigg\{ \Bigg[(1 + \tau)F^{2}_{A} - (1 - \tau)(F^{V}_{1})^{2} \nonumber \\
&&+ ~\tau(1 - \tau)(F^{V}_{2})^{2} + 4\tau F^{V}_{1}F^{V}_{2}\Bigg] \nonumber \\
&&- ~\frac{m^{2}_{l}}{4M^{2}_{N}}\Bigg[(F^{V}_{1} + F^{V}_{2})^{2} \nonumber \\
&&+ ~(F_{A} + 2F_{P})^{2} - 4(1 + \tau)F^{2}_{P}\Bigg]\Bigg\} ,   \\
B(Q^{2})&=&\frac{Q^{2}}{M^{2}_{N}} ~F_{A} ~(F^{V}_{1} + F^{V}_{2}) ,   \\
C(Q^{2})&=&\frac{1}{4} ~\Bigg[F^{2}_{A} + (F^{V}_{1})^{2} + \tau(F^{V}_{2})^{2}\Bigg] ,
\end{eqnarray}
where $\tau = \frac{Q^{2}}{4M^{2}_{N}}$. $F_{A}$ is the axial form factor, $F_{P}$ is the pseudoscalar form factor and $F^{V}_{1}$, $F^{V}_{2}$ are the vector form factors.

The axial form factor $F_{A}$ is defined in the dipole form as~\cite{Stoler:1993yk}:
\begin{equation}
F_{A}(Q^{2}) = \frac{g_{A}}{(1 + \frac{Q^{2}}{M^{2}_{A}})^{2}} ,
\end{equation}
where $g_{A} ~(= -1.267)$ is the axial vector constant and $M_{A}$ is the axial mass.

The pseudoscalar form factor $F_{P}$ is defined in terms of axial form factor $F_{A}$ as~\cite{Bernard:2001rs}: 
\begin{equation}
F_{P}(Q^{2}) = \frac{2 ~M^{2}_{N}}{Q^{2} + m^{2}_{\pi}} ~F_{A}(Q^{2}) ,
\end{equation}
where $m_{\pi}$ is the mass of pion.

The vector form factors $F^{V}_{1}$ and $F^{V}_{2}$ are defined as~\cite{Stoler:1993yk,Budd:2003wb}: 
\begin{eqnarray}
F^{V}_{1}(Q^{2})&=&\frac{1}{(1 + \tau)} \Bigg\{\Bigg[G^{p}_{E}(Q^{2}) - G^{n}_{E}(Q^{2})\Bigg] \nonumber \\
&&+ ~\tau \Bigg[G^{p}_{M}(Q^{2}) - G^{n}_{M}(Q^{2})\Bigg]\Bigg\},  \\
F^{V}_{2}(Q^{2})&=&\frac{1}{(1 + \tau)} \Bigg\{\Bigg[G^{p}_{M}(Q^{2}) - G^{n}_{M}(Q^{2})\Bigg] \nonumber \\
&&- \Bigg[G^{p}_{E}(Q^{2}) - G^{n}_{E}(Q^{2})\Bigg]\Bigg\} ,
\end{eqnarray}
where $G^{p}_{E}$ is the electric Sach's form factor of proton, $G^{n}_{E}$ is the electric Sach's form factor of neutron, $G^{p}_{M}$ is the magnetic Sach's form factor of proton and $G^{n}_{M}$ is the magnetic Sach's form factor of neutron. Several groups such as Galster et al.~\cite{Galster:1971kv}, Budd et al.~\cite{Budd:2004bp}, Bradford et al.~\cite{Bradford:2006yz}, Bosted~\cite{Bosted:1994tm} and Alberico et al.~\cite{Alberico:2008sz} provide parameterizations of these form factors by fitting the electron scattering data. For present calculations, we are using Galster et al. parameterizations of these form factors.

The electric and magnetic Sach's form factors of nucleons are defined as~\cite{Galster:1971kv}:
\begin{eqnarray}
G^{p}_{E}(Q^{2})&=&G_{D}(Q^{2}) ,  \\
G^{p}_{M}(Q^{2})&=&\mu_{p} ~G_{D}(Q^{2}) ,  \\
G^{n}_{M}(Q^{2})&=&\mu_{n} ~G_{D}(Q^{2}) . 
\end{eqnarray}
We define the electric Sach's form factor of neutron using Krutov and Troitsky~\cite{Krutov:2002tp} parameterization as:
\begin{equation}
G^{n}_{E}(Q^{2}) = -\mu_{n} ~\frac{0.942 ~\tau}{(1 + 4.61 ~\tau)} G_{D}(Q^{2}) ,
\end{equation}
where $\mu_{p} ~(= 2.793)$ is the magnetic moment of proton, $\mu_{n} ~(= -1.913)$ is the magnetic moment of neutron and $G_{D}(Q^{2})$ is the dipole form factor defined as~\cite{Stoler:1993yk}:
\begin{equation}
G_{D}(Q^{2}) = \frac{1}{\Bigg(1 + \frac{Q^{2}}{M^{2}_{v}}\Bigg)^{2}} ,
\end{equation}
where $M^{2}_{v} = 0.71$ GeV$^{2}$.

The total cross section of anti-neutrino - nucleon (free) quasi elastic scattering is obtained by integrating the differential cross section defined by Eq. \ref{eq:dsigma_free_nucleon} over $Q^{2}$ as~\cite{TotalXSectionLabel}:
\begin{equation}
\sigma^{free}(E_{\bar{\nu}}) = \int^{Q^{2}_{max}}_{Q^{2}_{min}} dQ^{2} ~\frac{d\sigma^{free}(E_{\bar{\nu}},Q^{2})}{dQ^{2}} ,
\end{equation}
where $Q^{2}_{min}$ and $Q^{2}_{max}$ are defined as:
\begin{eqnarray}
Q^{2}_{min}&=& - m^{2}_{l} + 2 ~E_{\bar{\nu}} ~(E_{l} - |\vec{k'}|) ,     \nonumber \\   
&=& \frac{2 E^{2}_{\bar{\nu}}M_{N} - M_{N} m^{2}_{l} - E_{\bar{\nu}}m^{2}_{l} - E_{Q}}{2 E_{\bar{\nu}} + M_{N}} .  \label{eq:Q2min}      \\
Q^{2}_{max}&=& - m^{2}_{l} + 2 ~E_{\bar{\nu}} ~(E_{l} + |\vec{k'}|) ,     \nonumber \\   
&=& \frac{2 E^{2}_{\bar{\nu}}M_{N} - M_{N} m^{2}_{l} + E_{\bar{\nu}}m^{2}_{l} + E_{Q}}{2 E_{\bar{\nu}} + M_{N}} .  \label{eq:Q2max}
\end{eqnarray}
Here, $E_{l}$ and $\vec{k'}$ are the energy and momentum of the outgoing lepton and $E_{Q}$ is defined as:
\begin{equation}
E_{Q} = E_{\bar{\nu}}\sqrt{(s - m^{2}_{l})^{2} - 2(s + m^{2}_{l})M^{2}_{N} + M^{4}_{N}} ,
\end{equation}
where $s = M^{2}_{N} + 2M_{N}E_{\bar{\nu}}$.

\section{Nuclear modifications}

For studying anti-neutrino - nucleus quasi elastic scattering, nucleus can be treated as a Fermi gas~\cite{LlewellynSmith:1971uhs,Smith:1972xh,Kuzmin:2007kr}, where the nucleons move independently within the nuclear volume in an average binding potential generated by all nucleons. Pauli suppression condition is applied for the nuclear modifications which implies that the cross section for all the interactions leading to a final state nucleon with a momentum smaller than the Fermi momentum $k_{F}$ is equal to zero.

The differential cross section per proton for anti-neutrino - nucleus quasi elastic scattering is defined as: 
\begin{eqnarray}
\frac{d\sigma^{nucleus}(E_{\bar{\nu}},Q^{2})}{dQ^{2}}&=&\frac{2V}{Z(2\pi)^{3}} \int^{\infty}_{0} 2\pi k^{2}_{p}dk_{p}d(cos\theta) \nonumber \\
&&f(\vec{k_{p}})S(\nu - \nu_{min}) \nonumber \\
&&\frac{d\sigma^{free}(E^{eff}_{\bar{\nu}}(E_{\bar{\nu}},\vec{k_{p}}),Q^{2})}{dQ^{2}} ,  \label{eq:dsigma_nucleus}
\end{eqnarray}
where the factor 2 accounts for the spin of the proton, $V$ is the volume of the nucleus, $k_{p}$ is the momentum of the proton, $\frac{d\sigma^{free}}{dQ^{2}}$ is the differential cross section of the anti-neutrino quasi elastic scattering off free proton as defined by Eq. \ref{eq:dsigma_free_nucleon} and $E^{eff}_{\bar{\nu}}$ is the effective anti-neutrino energy in the presence of Fermi motion of nucleons.

$E^{eff}_{\bar{\nu}}$ is defined as: 
\begin{equation}
E^{eff}_{\bar{\nu}} = \frac{(s^{eff} - M^{2}_{p})}{2M_{p}} .
\end{equation}
Here, $M_{p}$ is the proton mass and $s^{eff}$ is defined as: 
\begin{equation}
s^{eff} = M^{2}_{p} + 2E_{\bar{\nu}}\Bigg(E_{p} - k_{p} cos\theta\Bigg) .
\end{equation}
where $E_{p}$ is the proton energy defined as: 
\begin{equation}
E_{p} = \sqrt{k^{2}_{p} + M^{2}_{p}} .
\end{equation}

The Fermi distribution function $f(\vec{k_{p}})$ is defined as: 
\begin{equation}
f(\vec{k_{p}}) = \frac{1}{1 + exp(\frac{k_{p} - k_{F}}{a})} ,
\end{equation}
where $a = kT ~(= 0.020$ GeV$)$ is the diffuseness parameter~\cite{Saraswat:2016tqa}. The Fermi momentum $k_{F}$ for carbon nucleus is $0.221$ GeV~\cite{Moniz:1971mt}.

The Pauli suppression factor $S(\nu - \nu_{min})$ is defined as: 
\begin{equation}
S(\nu - \nu_{min}) = \frac{1}{1 + exp(- ~\frac{(\nu - \nu_{min})}{a})} ,
\end{equation}
where $\nu$ is the energy transfer in the interaction defined as: 
\begin{equation}
\nu = (Q^{2} + M^{2}_{n} - M^{2}_{p}) / (2M_{p}) .
\end{equation}
and $\nu_{min}$ is defined as: 
\begin{equation}
\nu_{min} = \sqrt{k^{2}_{F} + M^{2}_{n}} - \sqrt{k^{2}_{p} + M^{2}_{p}} + E_{B} .
\end{equation}
Here, $M_{n}$ is the final state neutron mass and $E_{B}$ is the binding energy. For carbon nucleus, $E_{B} = 10$ MeV~\cite{Saraswat:2016tqa}.

The total cross section of anti-neutrino - nucleus quasi elastic scattering is obtained by integrating the differential cross section as defined by Eq. \ref{eq:dsigma_nucleus} over $Q^{2}$, where $Q^{2}$ ranges from $Q^{2}_{min}$ to $Q^{2}_{max}$ defined by Eqs. \ref{eq:Q2min} and \ref{eq:Q2max} calculated with $E^{eff}_{\bar{\nu}}$ instead of $E_{\bar{\nu}}$.

\section{Results and discussions}

We calculated the charged current $\bar{\nu} - N$ and $\bar{\nu} - A$ quasi elastic differential scattering cross sections. Fig. \ref{fig:dsigma_nucleon_2GeV} shows the present calculations of $\bar{\nu}_{\mu} - p$ charged current quasi elastic differential scattering cross section $\frac{d\sigma}{dQ^{2}}$ as a function of the square of momentum transfer $Q^{2}$, for different values of axial mass $(M_{A} = 0.979, ~1.05, ~1.12$ and ~$1.23$ GeV$)$ and for anti-neutrino energy $E_{\bar{\nu}} = 2$ GeV. The value of differential cross section increases with increase in the value of axial mass.
 
Fig. \ref{fig:dsigma_comparison_2GeV} shows the differential cross section $\frac{d\sigma}{dQ^{2}}$ for $\bar{\nu}_{\mu} - p$ and $\bar{\nu}_{\mu} - ^{12}$C charged current QES as a function of the square of momentum transfer $Q^{2}$, with axial mass $M_{A} = 1.05$ GeV and anti-neutrino energy $E_{\bar{\nu}} = 2$ GeV. The anti-neutrino - carbon cross section is lower than the anti-neutrino - proton cross section for smaller values of $Q^{2}$ due to nuclear effects. The cross sections gradually drop to zero with increase in the value of $Q^{2}$.
\begin{center}
\includegraphics[width=9cm]{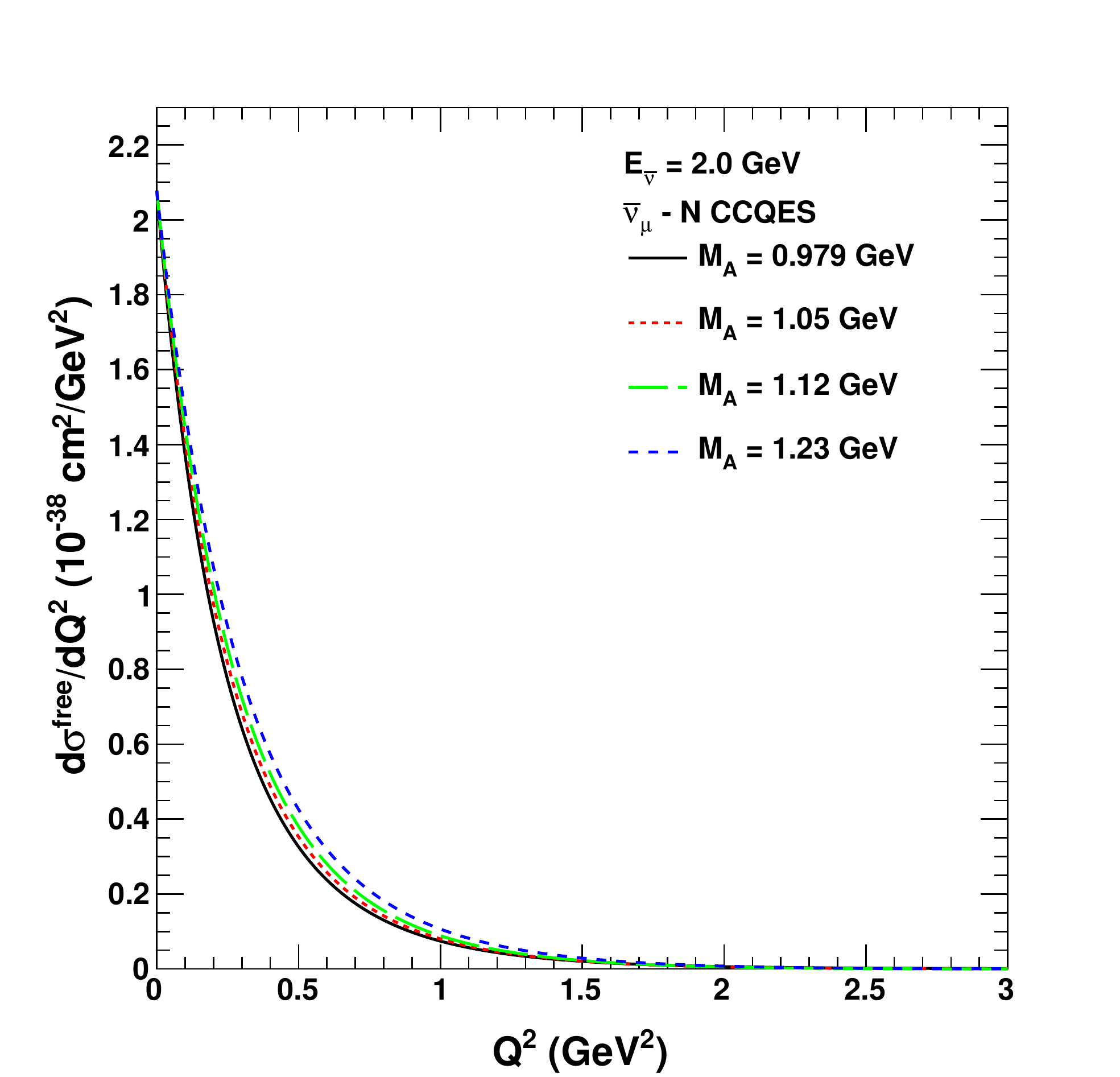}
\figcaption{\label{fig:dsigma_nucleon_2GeV} Differential cross section $\frac{d\sigma}{dQ^{2}}$ for $\bar{\nu}_{\mu} - p$ charged current QES as a function of the square of momentum transfer $Q^{2}$, for different values of axial mass $M_{A}$ and for anti-neutrino energy $E_{\bar{\nu}} = 2$ GeV.}
\end{center}
\begin{center}
\includegraphics[width=9cm]{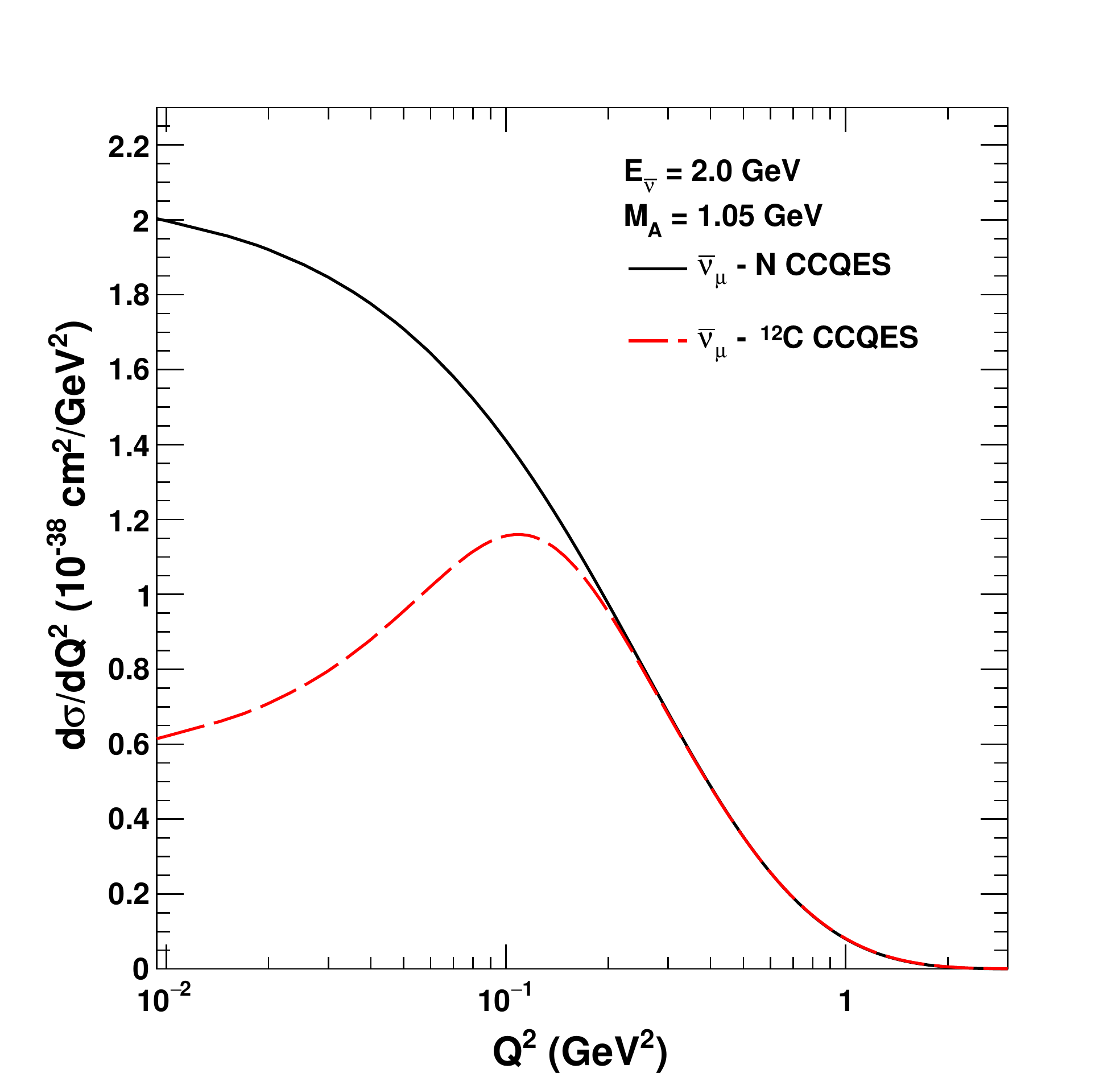}
\figcaption{\label{fig:dsigma_comparison_2GeV} Differential cross section $\frac{d\sigma}{dQ^{2}}$ for $\bar{\nu}_{\mu} - p$ and $\bar{\nu}_{\mu} - ^{12}$C charged current QES as a function of the square of momentum transfer $Q^{2}$, for axial mass $M_{A} = 1.05$ GeV and for anti-neutrino energy $E_{\bar{\nu}} = 2$ GeV.}
\end{center}

We compared the present calculations of $\bar{\nu}_{\mu} - ^{12}$C charged current quasi elastic differential scattering cross section with experimental data from several collaborations. Fig. \ref{fig:dsigma_nucleus_3_5GeV} shows the differential cross section $\frac{d\sigma}{dQ^{2}}$ per proton for anti-neutrino - carbon CCQES as a function of the square of momentum transfer $Q^{2}$, for different values of axial mass $(M_{A} = 0.979, ~1.05, ~1.12$ and ~$1.23$ GeV$)$. The results obtained are compared with MINER$\nu$A data~\cite{Fields:2013zhk} measuring muon anti-neutrino quasi elastic scattering on a hydrocarbon target at $<E_{\bar{\nu}}>$ = 3.5 GeV. The calculation with axial mass $M_{A} = 0.979$ GeV is compatible with data.
\begin{center}
\includegraphics[width=9cm]{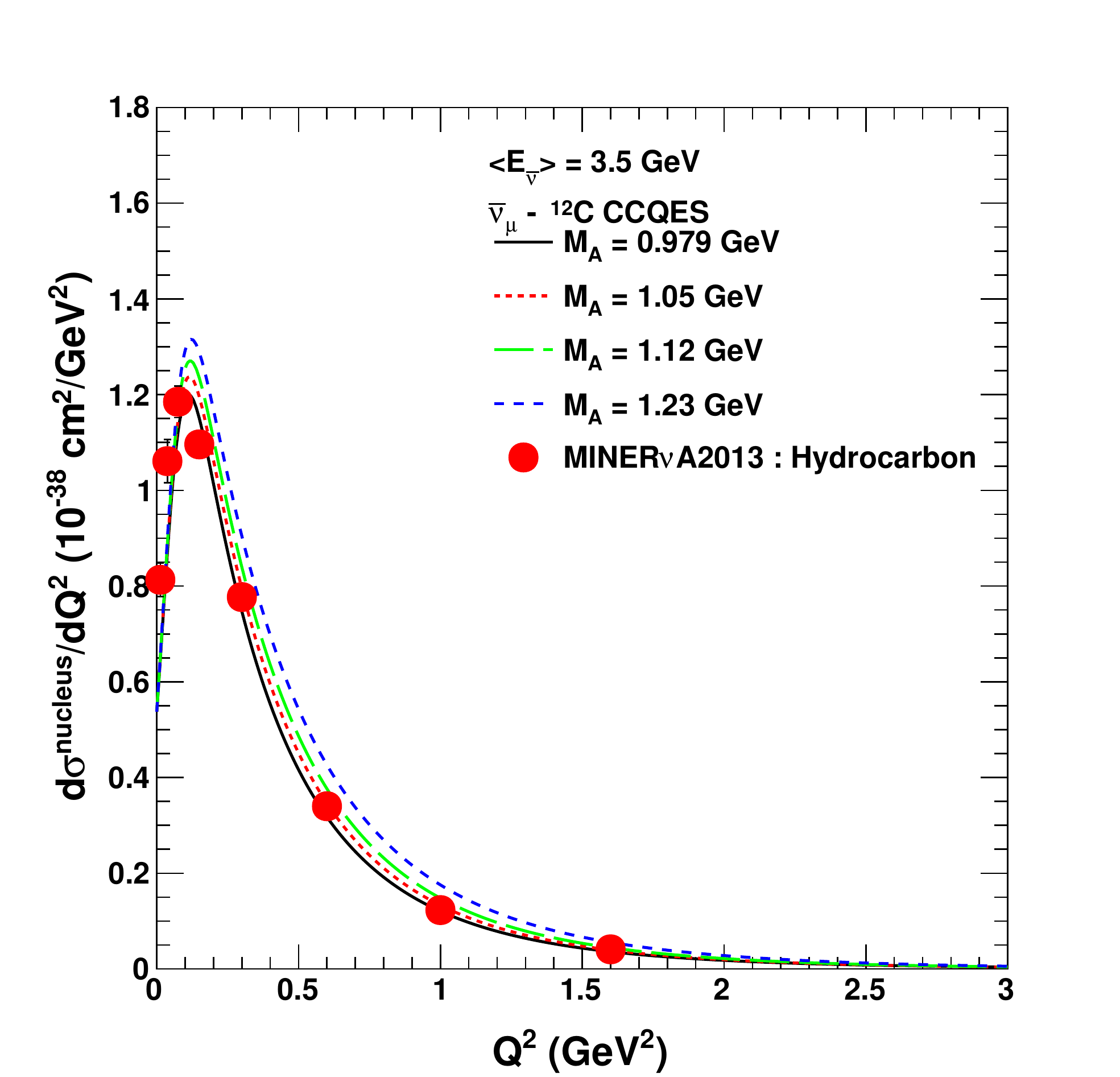}
\figcaption{\label{fig:dsigma_nucleus_3_5GeV} Differential cross section $\frac{d\sigma}{dQ^{2}}$ per proton for $\bar{\nu}_{\mu} - ^{12}$C charged current QES as a function of the square of momentum transfer $Q^{2}$, for different values of axial mass $M_{A}$ and for average anti-neutrino energy $<E_{\bar{\nu}}>$ = 3.5 GeV compared with MINER$\nu$A data~\cite{Fields:2013zhk}.}
\end{center}

Fig. \ref{fig:dsigma_nucleus_2GeV} shows the differential cross section $\frac{d\sigma}{dQ^{2}}$ per proton for anti-neutrino - carbon CCQES as a function of the square of momentum transfer $Q^{2}$, for different values of axial mass $(M_{A} = 0.979, ~1.05, ~1.12$ and ~$1.23$ GeV$)$ and for average anti-neutrino energy $<E_{\bar{\nu}}>$ = 2 GeV. The results obtained are compared with data from Gargamelle (GGM)~\cite{Bonetti:1977cs} studying quasi elastic reactions of neutrinos and antineutrinos on propane plus freon target. The calculations with axial mass $M_{A} = 0.979$ and $1.05$ GeV are compatible with data.
\begin{center}
\includegraphics[width=9cm]{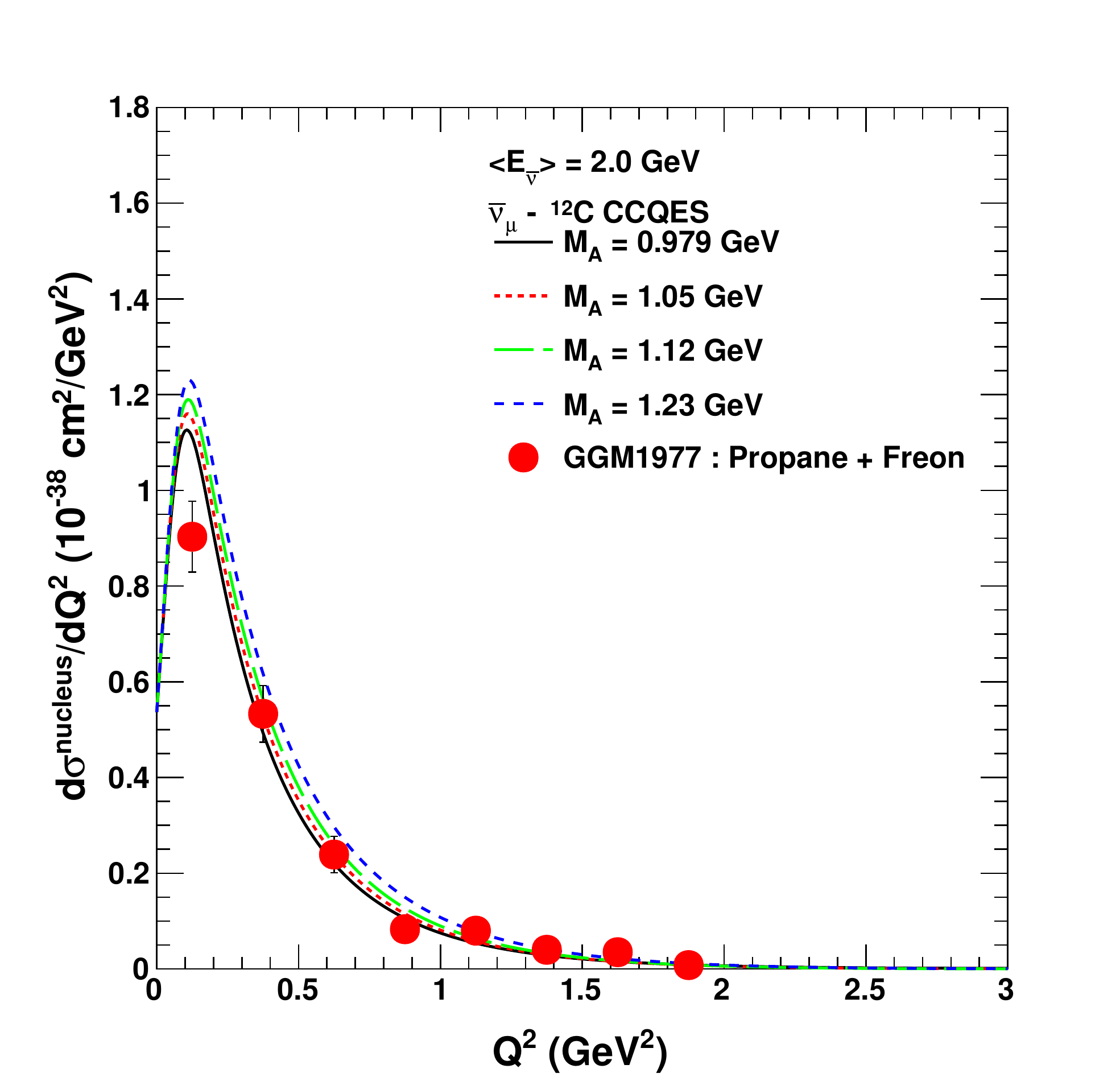}
\figcaption{\label{fig:dsigma_nucleus_2GeV} Differential cross section $\frac{d\sigma}{dQ^{2}}$ per proton for $\bar{\nu}_{\mu} - ^{12}$C charged current QES as a function of the square of momentum transfer $Q^{2}$, for different values of axial mass $M_{A}$ and for average anti-neutrino energy $<E_{\bar{\nu}}>$ = 2 GeV compared with GGM data~\cite{Bonetti:1977cs}.}
\end{center}

Fig. \ref{fig:dsigma_nucleus_SKATGeV} shows the differential cross section $\frac{d\sigma}{dQ^{2}}$ per proton for anti-neutrino - carbon CCQES as a function of the square of momentum transfer $Q^{2}$, for different values of axial mass $(M_{A} = 0.979, ~1.05, ~1.12$ and ~$1.23$ GeV$)$ and for average anti-neutrino energy $<E_{\bar{\nu}}>$ = 3 GeV. The results obtained are compared with SKAT data~\cite{Brunner:1989kw} studying the cross sections of neutrino and anti-neutrino quasi elastic interactions using a wide energy band (3 - 30 GeV) neutrino/anti-neutrino beam on heavy freon (CF$_{3}$Br) target. The calculations with axial mass $M_{A} = 0.979$ and $1.05$ GeV are compatible with data.

Fig. \ref{fig:dsigma_nucleus_MinibooneGeV} shows flux-integrated differential cross section $\frac{d\sigma}{dQ^{2}}$ per proton for anti-neutrino - carbon CCQES as a function of the square of momentum transfer $Q^{2}$ corresponding to the MiniBooNE data~\cite{Aguilar-Arevalo:2013dva}, measuring the muon anti-neutrino CCQES cross section off mineral oil (carbon) target. The calculations are performed for different values of axial mass $(M_{A} = 0.979, ~1.05, ~1.12$ and ~$1.23$ GeV$)$. The average anti-neutrino energy $<E_{\bar{\nu}}>$ = 665 MeV. The calculations with axial mass $M_{A} = 0.979$ and $1.05$ GeV are compatible with data.
\begin{center}
\includegraphics[width=9cm]{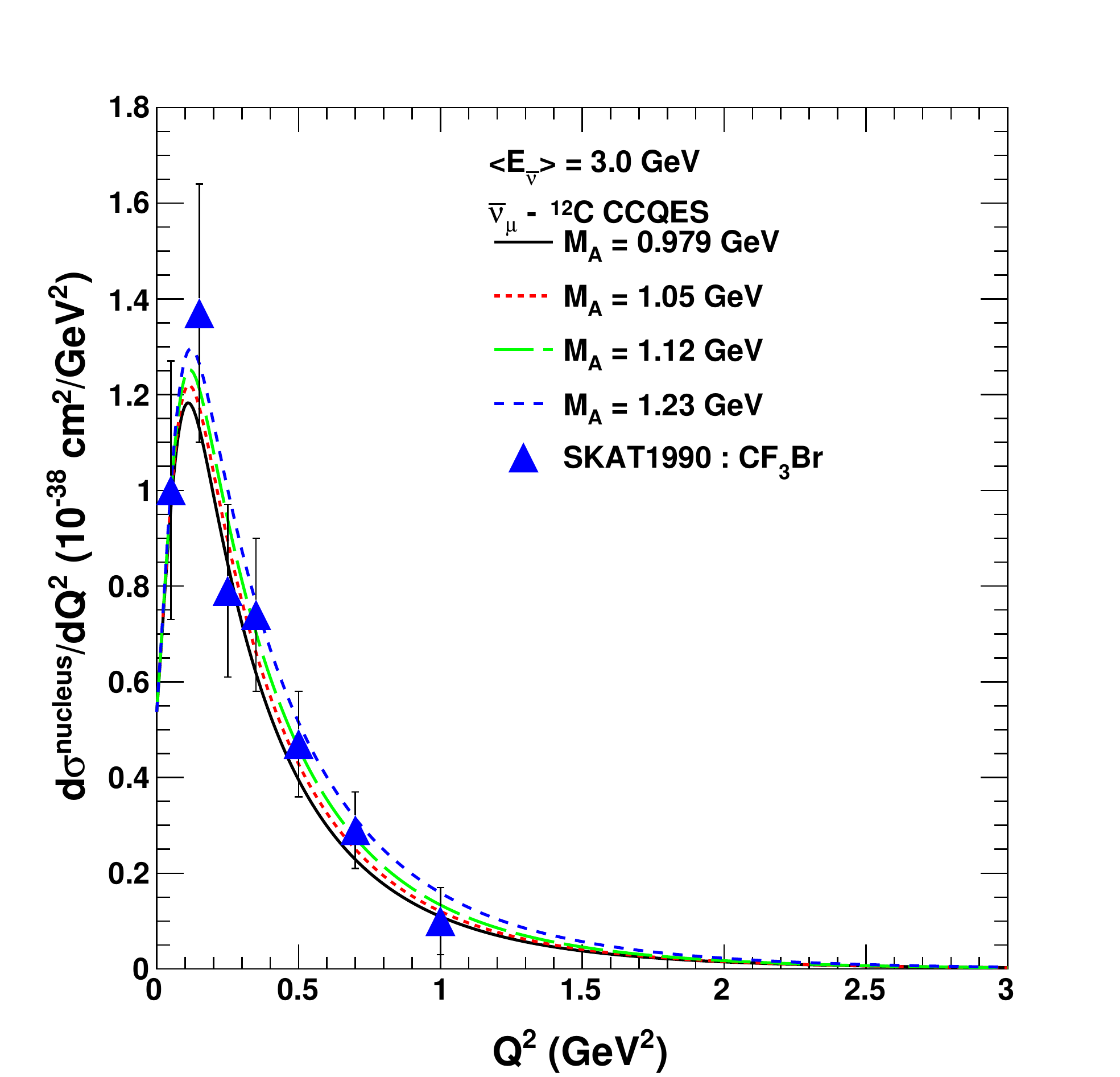}
\figcaption{\label{fig:dsigma_nucleus_SKATGeV} Differential cross section $\frac{d\sigma}{dQ^{2}}$ per proton for $\bar{\nu}_{\mu} - ^{12}$C charged current QES as a function of the square of momentum transfer $Q^{2}$, for different values of axial mass $M_{A}$ and for average anti-neutrino energy $<E_{\bar{\nu}}>$ = 3 GeV compared with SKAT data~\cite{Brunner:1989kw}.}
\end{center}
\begin{center}
\includegraphics[width=9cm]{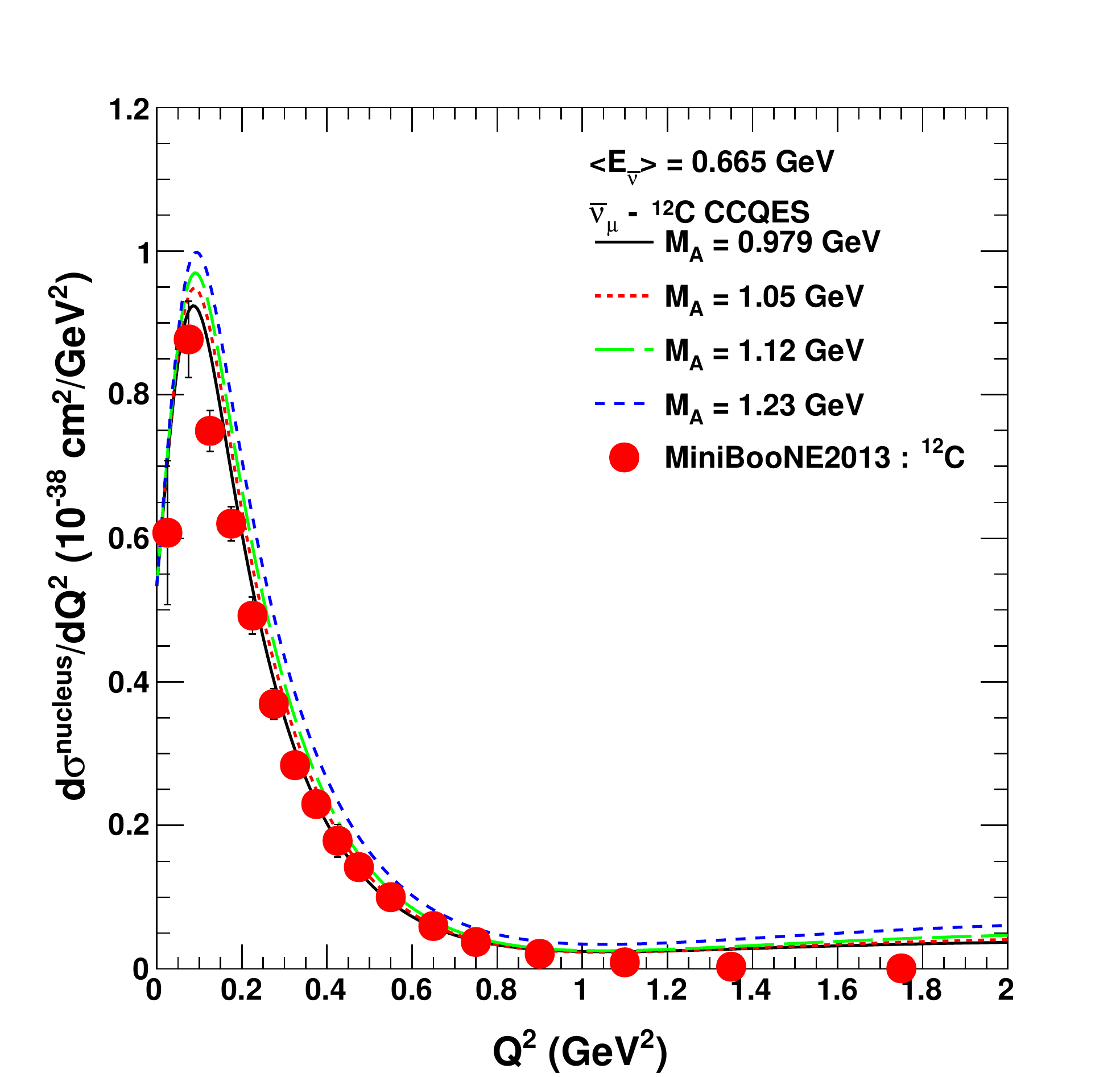}
\figcaption{\label{fig:dsigma_nucleus_MinibooneGeV} Flux-integrated differential cross section $\frac{d\sigma}{dQ^{2}}$ per proton for $\bar{\nu}_{\mu} - ^{12}$C charged current QES as a function of the square of momentum transfer $Q^{2}$ corresponding to the MiniBooNE data~\cite{Aguilar-Arevalo:2013dva}. The average anti-neutrino energy $<E_{\bar{\nu}}>$ = 665 MeV.}
\end{center}

We performed the calculations of the total cross section for charged current $\bar{\nu} - N$ and $\bar{\nu} - A$ quasi elastic scattering and compared the present results with data from several experiments. Fig. \ref{fig:total_sigma_nucleon} shows the present calculations of the total cross section $\sigma$ for anti-neutrino - proton CCQES as a function of the anti-neutrino energy $E_{\bar{\nu}}$, for different values of axial mass $(M_{A} = 0.979, ~1.05, ~1.12$ and ~$1.23$ GeV$)$. The value of total cross section increases with increase in the value of axial mass. We compared the obtained results with data from BNL~\cite{Fanourakis:1980si} and NOMAD~\cite{Lyubushkin:2008pe} experiments. The calculation with axial mass $M_{A} = 1.05$ GeV is compatible with data.

Fig. \ref{fig:total_sigma_comparison} shows the total cross section $\sigma$ for $\bar{\nu}_{\mu} - p$ and $\bar{\nu}_{\mu} - ^{12}$C charged current QES as a function of the anti-neutrino energy $E_{\bar{\nu}}$, with axial mass $M_{A} = 1.05$ GeV. The nuclear effects reduce anti-neutrino - carbon cross section compared to the anti-neutrino - proton cross section.
\begin{center}
\includegraphics[width=9cm]{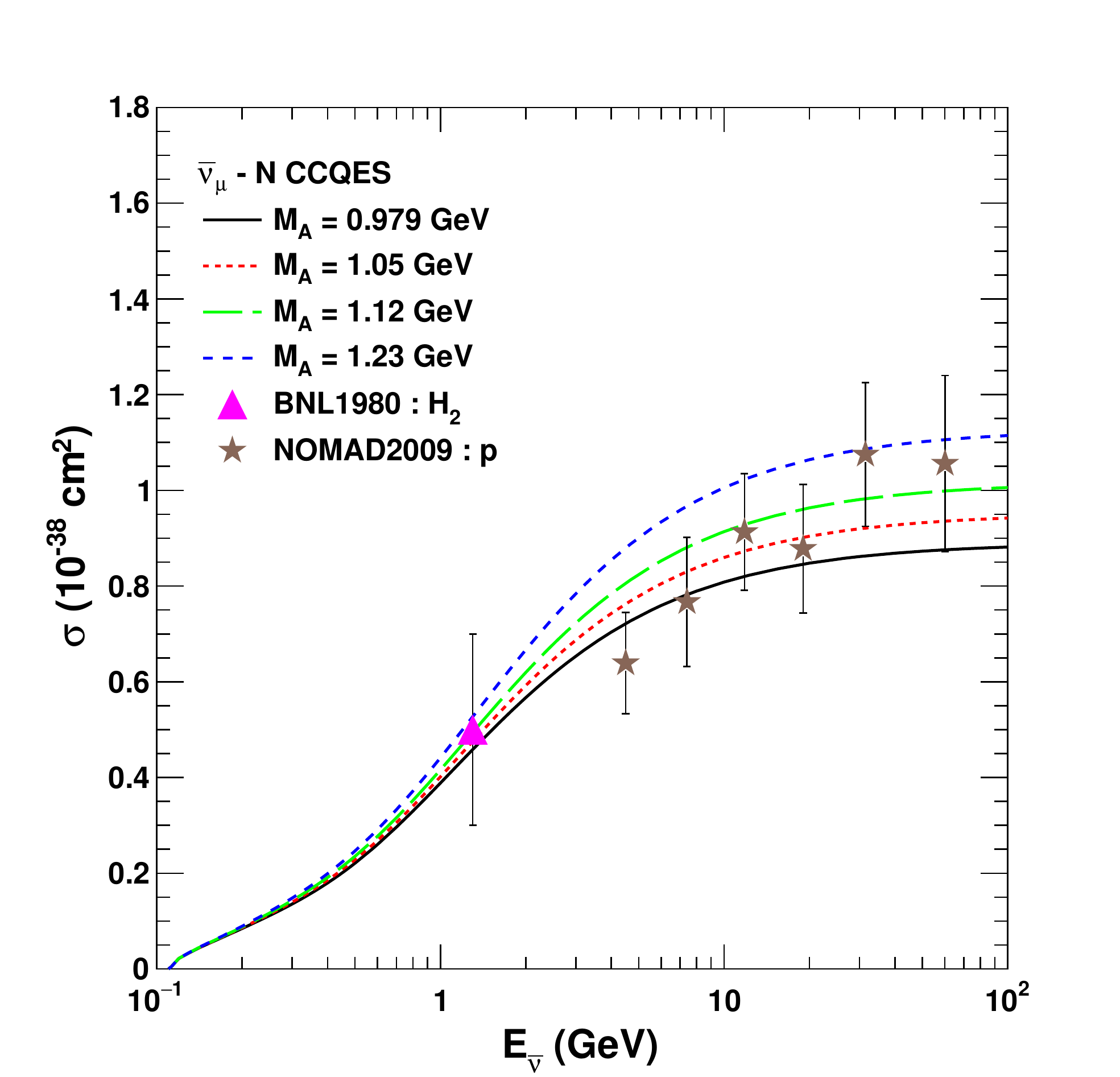}
\figcaption{\label{fig:total_sigma_nucleon} Total cross section $\sigma$ for $\bar{\nu}_{\mu} - p$ CCQES as a function of the anti-neutrino energy $E_{\bar{\nu}}$, for different values of axial mass $M_{A}$ compared with BNL~\cite{Fanourakis:1980si} and NOMAD~\cite{Lyubushkin:2008pe} data.}
\end{center}
\begin{center}
\includegraphics[width=9cm]{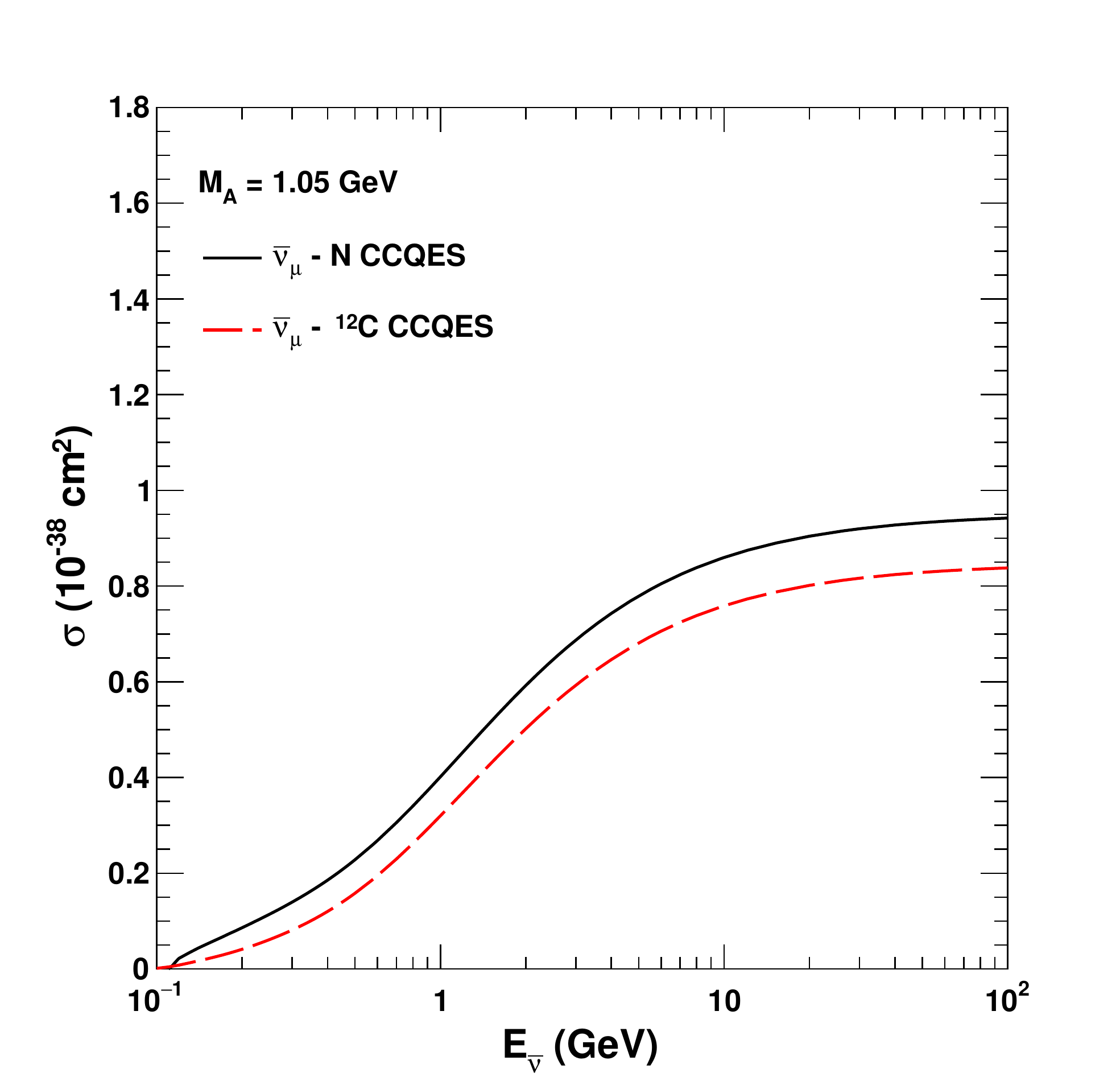}
\figcaption{\label{fig:total_sigma_comparison} Total cross section $\sigma$ for $\bar{\nu}_{\mu} - p$ and $\bar{\nu}_{\mu} - ^{12}$C charged current quasi elastic scattering as a function of the anti-neutrino energy $E_{\bar{\nu}}$, for axial mass $M_{A} = 1.05$ GeV.}
\end{center}

\begin{center}
\includegraphics[width=9cm]{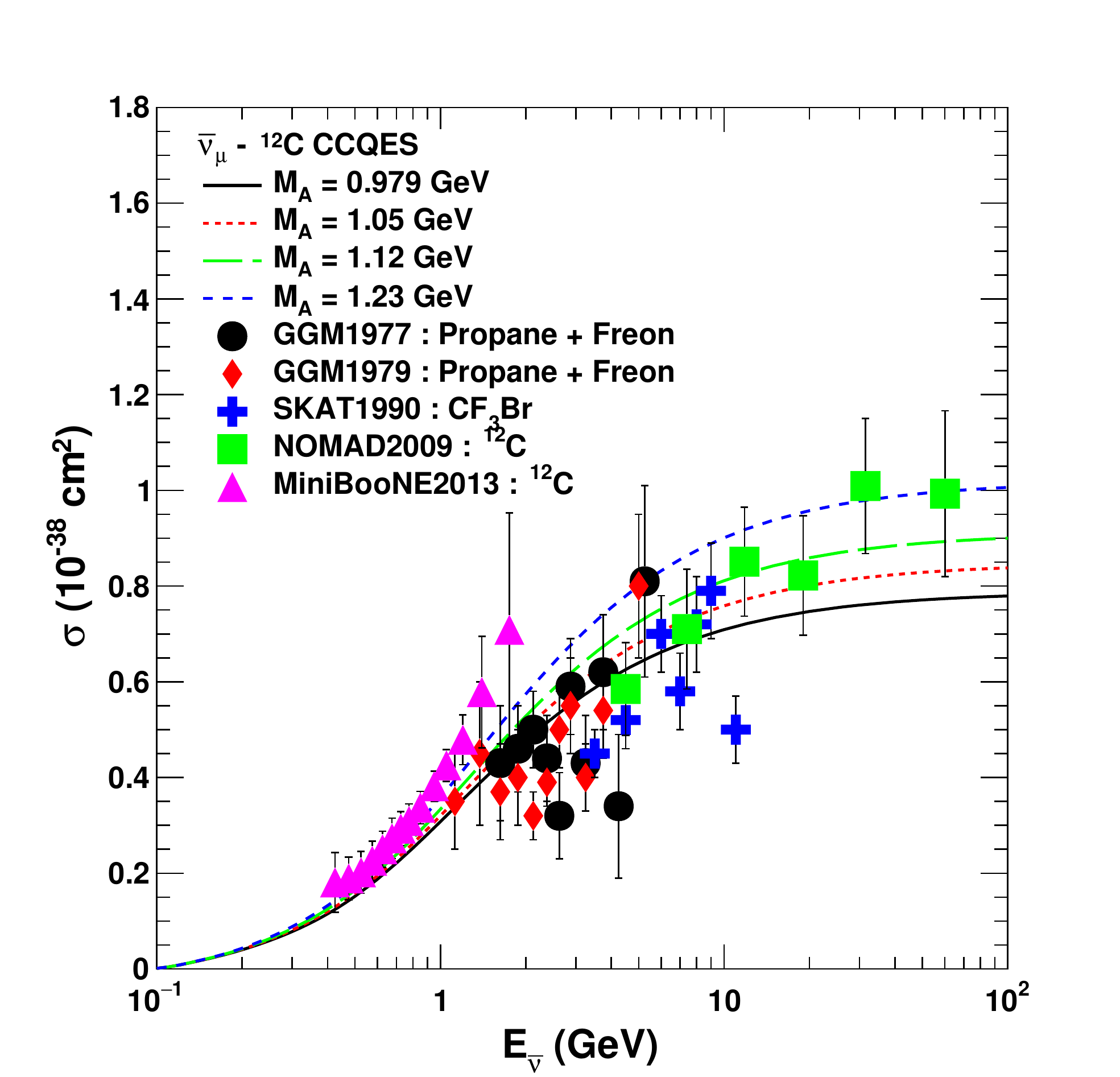}
\figcaption{\label{fig:total_sigma_nucleus} Total cross section $\sigma$ per proton for $\bar{\nu}_{\mu} - ^{12}$C charged current QES as a function of the anti-neutrino energy $E_{\bar{\nu}}$, for different values of axial mass $M_{A}$ compared with GGM(1977)~\cite{Bonetti:1977cs}, GGM(1979)~\cite{Armenise:1979zg}, SKAT~\cite{Brunner:1989kw}, NOMAD~\cite{Lyubushkin:2008pe} and MiniBooNE~\cite{Aguilar-Arevalo:2013dva} data.}
\end{center}

Fig. \ref{fig:total_sigma_nucleus} shows the total cross section $\sigma$ per proton for $\bar{\nu}_{\mu} - ^{12}$C charged current QES as a function of the anti-neutrino energy $E_{\bar{\nu}}$, for different values of axial mass $(M_{A} = 0.979, ~1.05, ~1.12$ and ~$1.23$ GeV$)$. The results obtained are compared with data from GGM(1977)~\cite{Bonetti:1977cs}, GGM(1979)~\cite{Armenise:1979zg}, SKAT~\cite{Brunner:1989kw}, NOMAD~\cite{Lyubushkin:2008pe} and MiniBooNE~\cite{Aguilar-Arevalo:2013dva} experiments. The calculations with axial mass $M_{A} = 0.979$ and $1.05$ GeV are compatible with GGM(1977), GGM(1979) and SKAT data though the calculations overestimate the data at low anti-neutrino energies. The approach parameterizing axial mass $M_{A}$ as a function of anti-neutrino energy, presented in Ref.~\cite{Kolupaeva:2016bfg}, can be used to get a better agreement with data at low anti-neutrino energies. The calculation with axial mass $M_{A} = 1.05$ GeV is compatible with NOMAD data and the calculation with axial mass $M_{A} = 1.23$ GeV is compatible with MiniBooNE data.

\section{Conclusion}

We presented a study on charged current anti-neutrino - nucleon and anti-neutrino - nucleus (carbon) quasi elastic scattering using Llewellyn Smith (LS) model. For electric and magnetic Sach's form factors of nucleons, we used Galster et al. parameterizations. Fermi gas model along with Pauli suppression condition has been used to incorporate the nuclear effects in anti-neutrino - nucleus QES. We calculated $\bar{\nu}_{\mu} - p$ and $\bar{\nu}_{\mu} - ^{12}$C charged current quasi elastic differential and total scattering cross sections for different values of axial mass $M_{A}$ and compared the obtained results with data from GGM, SKAT, BNL, NOMAD, MINER$\nu$A and MiniBooNE experiments. The present theoretical approach gives an excellent description of differential cross section data. The calculations with axial mass $M_{A} = 0.979$ and $1.05$ GeV are compatible with data from most of the experiments.




\end{multicols}

\vspace{-1mm}
\centerline{\rule{80mm}{0.1pt}}
\vspace{2mm}

\begin{multicols}{2}

\end{multicols}

\clearpage

\end{document}